\documentclass[aps,prl,reprint,groupedaddress]{revtex4-1}

\usepackage{graphicx,hyperref,amsmath,epstopdf,natbib,textcomp}

\begin{document}


\title{Anisotropic Stark effect and electric-field noise suppression for phosphorus donor qubits in silicon}


\author{A. J. Sigillito}
\email[]{asigilli@princeton.edu}
\affiliation{Princeton University, Dept. of Electrical Engineering}

\author{A. M. Tyryshkin}
\affiliation{Princeton University, Dept. of Electrical Engineering}

\author{S. A. Lyon}
\affiliation{Princeton University, Dept. of Electrical Engineering}


\date{\today}

\begin{abstract}

We report the use of novel, capacitively terminated coplanar waveguide (CPW) resonators to measure the quadratic Stark shift of phosphorus donor qubits in Si. We confirm that valley repopulation leads to an anisotropic spin-orbit Stark shift depending on electric and magnetic field orientations relative to the Si crystal. By measuring the linear Stark effect, we estimate the effective electric field due to strain in our samples. We show that in the presence of this strain, electric-field sources of decoherence can be non-negligible. Using our measured values for the Stark shift, we predict magnetic fields for which the spin-orbit Stark effect cancels the hyperfine Stark effect, suppressing decoherence from electric-field noise. We discuss the limitations of these noise-suppression points due to random distributions of strain and propose a method for overcoming them.

\end{abstract}


\maketitle



Quantum computing architectures based on donor qubits in silicon\cite{kane1998,morton2011,hill2005} have generated a great deal of interest due to their long coherence times exceeding seconds in isotopically enriched $^{28}$Si\cite{tyryshkin2012, wolfowicz2013}, scalability\cite{desousa2004,hollenberg2006}, and their compatibility with fabrication techniques refined by the Si industry. In many donor-based architectures the mechanism for manipulating individual spins relies on the tuning of a donor in and out of resonance with a global microwave magnetic field. Donor tuning via the Stark shift of Sb donors \cite{bradbury2006} and As donors \cite{lo2014} in Si have been studied using electron spin resonance (ESR) techniques \cite{mims1974, bradbury2006}. Similarly, strong electroelastic tuning of the hyperfine interaction for P donors in Si has been demonstrated using electrically detected magnetic resonance\cite{dreher2011}. 

In the present work, we measure the Stark shift of phosphorus donors in Si using ESR with novel, capacitively terminated, coplanar waveguide (CPW) resonators. These high-sensitivity resonators allow us to measure small spin ensembles subjected to locally homogeneous electric fields. Our measurements resolve a previously predicted anisotropy in the spin-orbit Stark shift due to valley repopulation \cite{rahman2009,pica2014}. 

It is expected that electric-field noise can contribute substantially to decoherence in the presence of strain. By measuring the linear Stark effect, we estimate the effective electric field due to strain in our devices and the resulting sensitivity of the donors to electric-field noise. From our results we find magnetic fields and crystal orientations where the spin-orbit and hyperfine components of the Stark shift cancel, such that spins are protected from electric-field noise. These ``noise-suppression points" are important for near-surface donors and quantum devices incorporating electrostatic gates.

\begin{figure}
\includegraphics{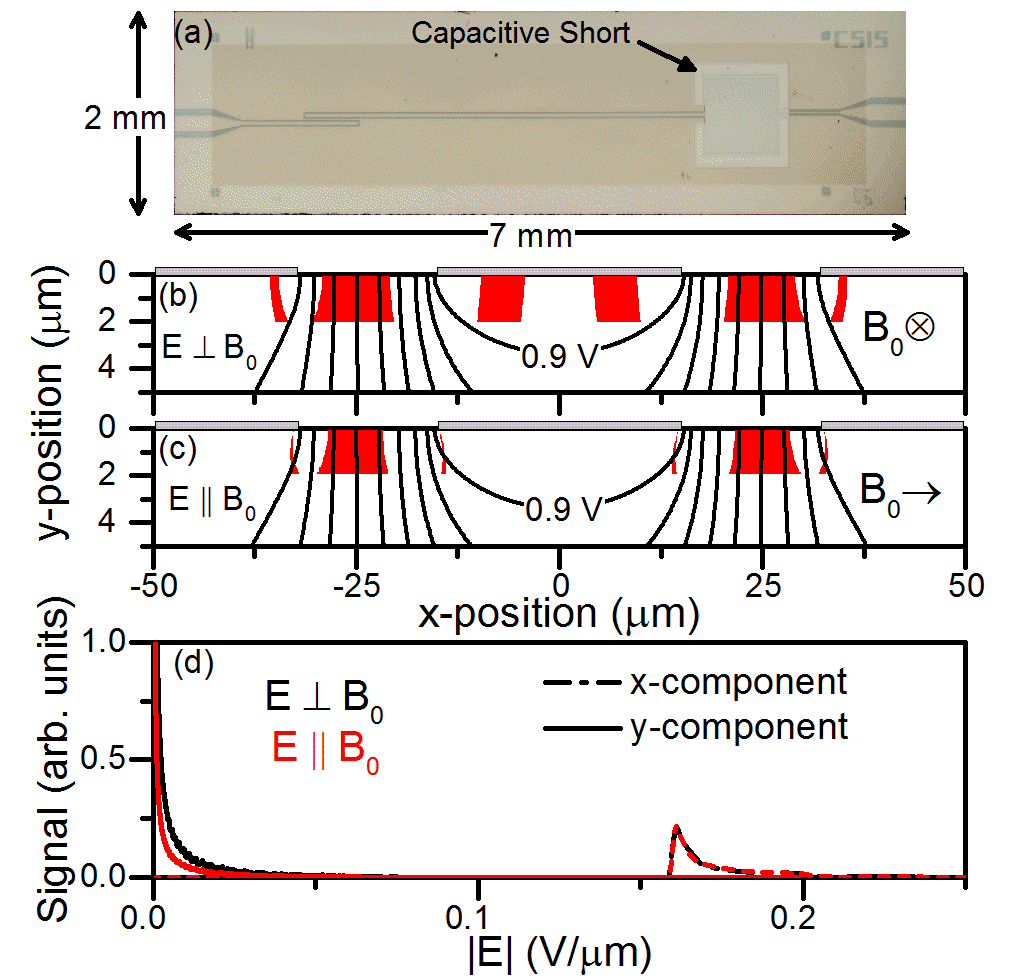}
\caption{\label{fig:fig1} (a) Optical micrograph of the CPW resonator. Microwaves excite the resonator from the transmission line on the left and a DC bias is applied through the wire on the right. Electric potential lines are shown at an antinode in $\vec{B}_{1}$ for $\vec{E} \perp\vec{B}_{0}$ (b) and $\vec{E}  \parallel \vec{B}_{0}$ (c). The CPW cross section is depicted by a cartoon at the top of the plot. The $0.9 \ V$ line is labeled (given a $1 \ V$ bias) and each subsequent line represents a $0.1 \ V$ decrease in potential. Red shaded regions denote where 50\% of the ESR signal originates. Microwave power was optimized to enhance sensitivity to spins at the center of the CPW gap where $\vec{E}$ is most uniform. (d) Electric field distribution in the CPW (given a $4 \ V$ bias) weighted by the signal contribution.} 
\end{figure}

Our experiments were conducted using 1/4-wavelength superconducting CPW resonators operating at 7.1 GHz with Q-factors of $\sim$1200. An example is shown in Fig.~\ref{fig:fig1}(a). Resonators were patterned from 35 nm thick Nb films deposited on 2 $\mu$m, P doped, $^{28}$Si epilayers (800 ppm $^{29}$Si). One end of a resonator is capacitively coupled to a single port transmission line used for exciting the resonator and measuring the spin signal. The other end is capacitively shorted to ground. This capacitive short consists of a 2.9 nF parallel plate capacitor having a plate area of 0.5 \ mm$^{2}$. The capacitor is filled with a 17 nm thick atomic layer deposition grown Al$_{2}$O$_{3}$ dielectric and satisfies the design rule for a good capacitive short \cite{simons2001}, $ 2 \pi f C Z_{0} \geq {50}$, where $f$ is the resonator frequency, $C$ is capacitance, and $Z_{0}$ is the characteristic impedance of the CPW (50 $\Omega$).

The capacitive short allows the center conductor of the CPW to be biased, providing a direct current (DC) electric field between the center pin and the ground plane of the resonator as shown in Fig.~\ref{fig:fig1}(b)-(c). This electric field ($\vec{E}$) is inhomogeneous, except near the plane of the CPW and in the gap between the center pin and the ground plane. To confine spins to these homogeneous regions, we employed a thin, 2 $\mu$m, phosphorus-doped $^{28}$Si epitaxial layer grown on high resistivity p-type Si as our spin ensemble.  However, the regions producing the ESR signal vary depending on sample orientation because only the microwave magnetic field ($\vec{B}_{1}$) perpendicular to the DC magnetic field ($\vec{B}_{0}$) drives spin rotations. The relevant components of $\vec{B_{1}}$ are plotted in the supplementary information. We calculated the sub-ensembles of spins contributing to at least half of the signal and they are shaded red in Fig.~\ref{fig:fig1}(b)-(c). The weighted electric-field distribution over the subensemble in the $\vec{E}\perp \vec{B}_{0}$ case is shown in Fig.~\ref{fig:fig1}(d). The standard deviation of the distribution in the x-component magnitude (the dominant component which is directed from the ground plane to the center pin), $|E_{x}|$, is 8\% and is approximately the same for $\vec{E} \parallel  \vec{B}_{0}$.

Two sets of resonators were fabricated with center pins oriented parallel to either the $[$010$]$ or the $[\overline{1}10]$ crystal axes. These orientations place $\vec{E}$ along either the $[$100$]$ or $[$110$]$ axes for spins contributing to the ESR signal. The resonators were wire bonded to copper printed circuit boards, connected to a low-noise, cryogenic preamplifier, and placed in a DC magnetic field. With $\vec{B}_{0}$ oriented in the plane of the Nb to avoid trapping magnetic flux vortices, devices were cooled to 1.7 K to conduct pulsed ESR experiments.


A pulsed ESR technique sensitive to small resonance shifts \cite{bradbury2006, mims1974} was used to measure the quadratic Stark shift. This technique uses a two-pulse Hahn echo sequence ($\pi/2(x) - \tau - \pi(y) - \tau - echo $), with an electric-field pulse applied to the spins during the first dephasing period, $\tau$. The electric field detunes the spins relative to the driving microwaves such that they accumulate an additional phase, $d\phi$, in the Hahn echo which is measured using a quadrature detector. This experiment utilized bipolar electric-field pulses (pulse sequence IV in \cite{bradbury2006}) consisting of a positive pulse immediately followed by a negative pulse of the same amplitude and duration. These pulses refocus linear Stark effects (arising from random strain as discussed below) thus allowing the measurement of the quadratic Stark effect. The phase shift is given by 
\begin{equation}
d\phi =\tau df= [\eta_{g} g \beta B_{0} + \eta_{a} a M_{I}]\vec{E}^{2}\tau_{E}/\hbar
\end{equation} where $df$ is the frequency shift of the spins, $\eta_{g}$ and $\eta_{a}$ are the spin-orbit and hyperfine Stark fitting parameters, respectively, $g$ is the electron g-factor, $\beta$ is the Bohr magneton, $a$ is the hyperfine coupling constant, $M_{I}$ is the nuclear spin projection, $\tau_{E}$ is the duration of the electric-field pulse, and $\hbar$ is the reduced Planck constant \cite{bradbury2006}. A model was developed to simulate the Stark effect in our device including the inhomogeneity in both $\vec{B_{1}}$ and $\vec{E}$. Both fields were computed using a conformal mapping technique\cite{wen1969}, and each spin's contribution to the echo was calculated as described in \cite{sigillito2014}. This model also took into account the $\vec{B_{1}}$ inhomogeneity present along the length of the CPW. The total echo phase shift of the spin ensemble, $\Delta \phi$, was determined by taking a weighted average of the Stark shift of each spin:
\begin{equation}
Ae^{i \Delta \phi} = \frac{\sum\limits_{i} e^{-i d\phi_{i}} g_{i} sin^{3} (g \beta B_{1i} \tau_{p}/\hbar)}{\sum\limits_{i} g_{i} sin^{3} (g \beta B_{1i} \tau_{p}/\hbar)}
\end{equation}where the sum is taken over all $i$ spins, $A$ is an amplitude coefficient, $d\phi_{i}$ is the phase shift of the $i$th spin, $g_{i}$ is the coupling of the $i$th spin to the resonator, $B_{1i}$ is the microwave magnetic field seen by the $i$th spin, and $\tau_{p}$ is the duration of the first microwave pulse in the Hahn echo sequence. In this expression, the $sin^{3}$ term takes into account signal attenuation due to pulse errors arising from $\vec{B}_{1}$ inhomogeneity \cite{malissa2013}. We note that the phase shift of a single spin, $d\phi_{i},$ is not affected by these errors \cite{Schweiger2001}.

\begin{figure}
\includegraphics{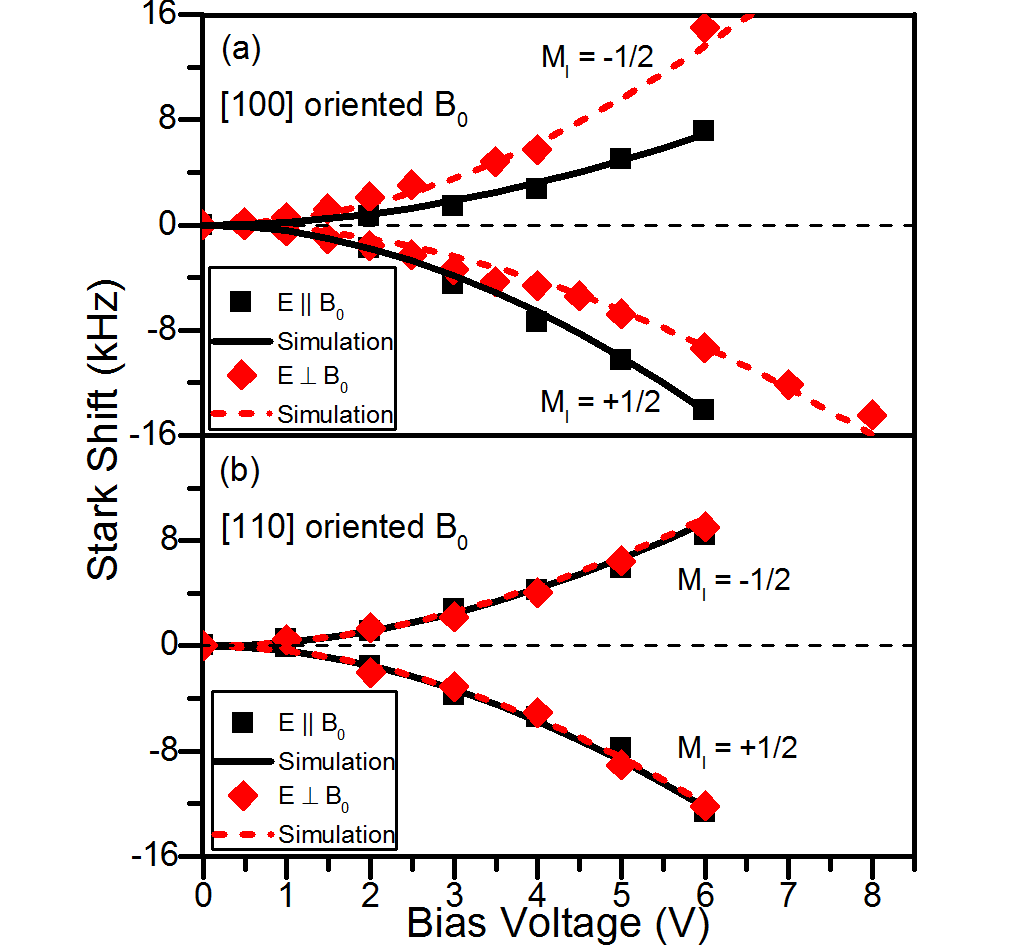}
\caption{\label{fig:fig2}Measured ESR frequency shift as a function of CPW center pin voltage for $\vec{B}_{0}$ oriented along the $[$100$]$ (a) and $[$110$]$ (b) crystal axes. Data are plotted for $\vec{E}$ applied either parallel (red diamond) or perpendicular (black square) to $\vec{B}_{0}$. The smooth curves are fits to the data  using Eq.(2). Note that the hyperfine Stark shift is symmetric whereas the spin-orbit Stark shift is asymmetric with respect to zero as a function of $M_{I}$. The error bars for the data are smaller than the points used in the plot. Data were taken at 1.7 K with a $B_{0}$ of $0.26 \ T$. } 
\end{figure}

Data for electric fields applied along the $[$100$]$ and $[$110$]$ axes with bias voltages up to 8 V are shown in Fig.~\ref{fig:fig2}. These data were taken with an electric-field pulse length of $38 \ \mu s$ and microwave $\pi$-pulses of $400 \ ns$. Eq.(2) was fitted to the data and the fitting parameters are given in Table 1. While the hyperfine Stark parameter remains nearly constant, the spin-orbit Stark parameter changes sign and magnitude depending on the electric field orientation relative to $\vec{B}_{0}$ and the crystal axis.

\begin{table*}
\caption{Stark Shift Fitting Parameters} 
\centering 
\begin{tabular}{c c c c c}
\hline\hline 
$\vec{E}$ Orientation  & $\vec{B}_{0}$ Orientation         & $\eta_{a}$($\mu m^{2}$/$V^{2}$)      & $\eta_{g}$($\mu m^{2}$/$V^{2}$)       & $\eta_{g} theory ^{\dagger}$ ($\mu m^{2}$/$V^{2}$) \\ [0.5ex]
\hline 
$[$100$]$ & $\vec{E} \ \parallel \ \vec{B}_{0}$ & $-2.6 \pm 0.1 \times 10^{-3}$ & $-8 \pm 2 \times 10^{-6}$  & $-12 \times 10^{-6}$  \\ 
$[$100$]$ & $\vec{E} \ \perp \ \vec{B}_{0}$ & $-2.8 \pm 0.1 \times 10^{-3}$ &$6 \pm 1.5 \times 10^{-6}$ & $14 \times 10^{-6}$  \\
$[$110$]$ & $\vec{E} \ \parallel \ \vec{B}_{0}$ & $-2.7 \pm 0.1 \times 10^{-3}$ &$-3.5 \pm 1.5 \times 10^{-6}$ & - \\
$[$110$]$ & $\vec{E} \ \perp \ \vec{B}_{0}$ & $-2.7 \pm 0.1 \times 10^{-3}$ & $-2.8 \pm 1.5 \times 10^{-6}$ & - \\ [1ex] 
\hline 
\end{tabular}
\label{table:fittingparams} 

Theoretical values are obtained from tight binding calculations reported by Rahman \textit{et al.} in \cite{rahman2009}.
\end{table*}

This anisotropic Stark shift was predicted by Rahman \textit{et al.} \cite{rahman2009} and is explained using a valley repopulation model \cite{feher1961}. In Si, the g-factor of an electron is related to its effective mass in the direction of $\vec{B}_{0}$. The non-spherical valleys produce a non-uniform effective mass, such that the g-factor of spins in a single valley depends on the angle $\vec{B}_{0}$ makes with that valley. In the unperturbed ground state with a symmetric valley combination, the g-factor is equally averaged over all valleys, and no g-factor anisotropy can be resolved. However, when an electric field is oriented along a valley axis ($\{$100$\}$ for Si), the valley degeneracy is partially lifted, and electrons preferentially fill valleys oriented along the electric field. This changes the effective g-factor and induces an anisotropy in the Stark parameter $\eta_{g}$. The anisotropy is most pronounced when the electric field is oriented along a valley axis. Moreover, application of an electric field in the \{111\} axes would result in no valley repopulation, since it makes the same angle with all valleys. This orientation is not accessible in our device geometry, but we were able to apply a $[$110$]$ oriented electric field, and, as expected, the spin-orbit anisotropy became small as shown in Table 1.

It is through the Stark shift that $\vec{E}$ noise can decohere spins and we note the existence of ESR transitions insensitive to this noise. From Eq.(1), we infer that decoherence from $\vec{E}$ noise can differ substantially for the two donor nuclear spin projections, $M_{I}$. Moreover, $\vec{B}_{0}$ can be tuned such that the first two terms cancel, leading to a transition insensitive to $\vec{E}$. Due to the dependence of $\eta_{g}$ on the orientation of $\vec{E}$, these noise-suppression points vary with the direction of $\vec{E}$, and no single $\vec{B}_{0}$ can cancel all randomly oriented electric-field noise. For this reason, noise suppression points are most effective in situations where the primary source of noise (and thus the direction of $\vec{E}$) is known, such as in gated donor architectures. 

In the simplest picture, electric-field noise (which is presumably small) should not contribute to decoherence, since the Stark effect is quadratic. However, this changes dramatically when a large DC electric field (present in gate-addressed architectures) or strain is applied. In this work we consider this field coming from strain \cite{bradbury2007aip}, inducing a shift in the ESR frequency by modulating the hyperfine interaction\cite{dreher2011} and causing valley repopulation\cite{feher1961}. The Stark shift in the presence of strain is given as
\begin{equation}
df = [\eta_{g} g \beta B_{0} + \eta_{a} a M_{I}](\vec{E}^{2}+2\vec{E} \cdot \vec{E}_{strain})/\hbar,
\end{equation} where $\vec{E}_{strain}$ is the effective electric field due to strain. When electric-field noise, $\vec{E}_{noise}$, is small, the $\vec{E}^{2}_{noise}$ term is negligible whereas the $\vec{E}_{noise} \cdot \vec{E}_{strain}$ term can be large. Using data in \cite{feher1961}, we calculate that a strain of $10^{-3}$ is equivalent to an effective electric field of $10 \ V/ \mu m$. We hereafter refer to strain in units of $V/ \mu m$, which corresponds to the strength of $\vec{E}_{strain}$.

To investigate strain we use unipolar (positive bias) pulses, instead of the bipolar pulses used to gather the data for Fig.~\ref{fig:fig2}. Bipolar pulses cancel the linear term of Eq.(3) whereas unipolar pulses lead to a broadening of the ESR line and loss of signal when internal strains are inhomogeneous across the spin ensemble.  Despite this broadening, data were taken using unipolar pulses with amplitudes of up to $4 \ V$ (supplemental material). These data indicate that strain is approximately $1 \ V/\mu m$ in our wafers, and the signal loss indicates that the strain is primarily random. We compare this to a similar sample (10 $\mu$m P doped epitaxial layer of $^{28}$Si) reported \cite{tyryshkin2006} with an electron nuclear double resonance (ENDOR) linewidth of 100 kHz. Assuming that broadening of the ENDOR line is due primarily to strain, this linewidth corresponds to an $\vec{E}_{strain}$ distribution spanning $0.75 \ V/\mu m$, comparable to our results. For these strains, the spins' sensitivity to electric-field noise, $df/dE$, can be large. The $\vec{E}_{noise}$ limited coherence time ($T_{2}$) is inversely proportional to $(df/dE)^{2}$ \cite{desousa2009} so electric-field sources of decoherence can be substantial. While the $\vec{E}_{noise}$ contribution to the coherence time will vary from system to system, we calculate (using \cite{desousa2009}) that for $10 \ \mu V$ of noise on our CPW center pin, $T_{2}$ would be limited to $70 \ ms$ in the devices used in these experiments Devices with gates more strongly coupled to the donors would have their $T_{2}$ affected more severely.

Internal strains can also lead to errors in measuring $\eta_{g}$, even when using bipolar pulses. This is because $\eta_{g}$ depends on the direction of the total electric field which includes $\vec{E}_{strain}$. Since $\vec{E}_{strain}$ is random, there is uncertainty in the actual direction of the total electric field relative to $\vec{B}_{0}$. Care was taken to minimize strain when mounting the devices, but not all strain can be avoided and in our devices we found that $\vec{E}_{strain}$ was larger than the applied electric fields. The errors arising from strain should become small as $\vec{E}$ becomes large, so data taken at higher bias voltages were weighted more heavily when fitting the model. For our samples we estimate that the magnitude of $\eta_{g}$ can be underestimated by up to a factor of 2.5 for the $[$100$]$ data.

\begin{figure}
\includegraphics{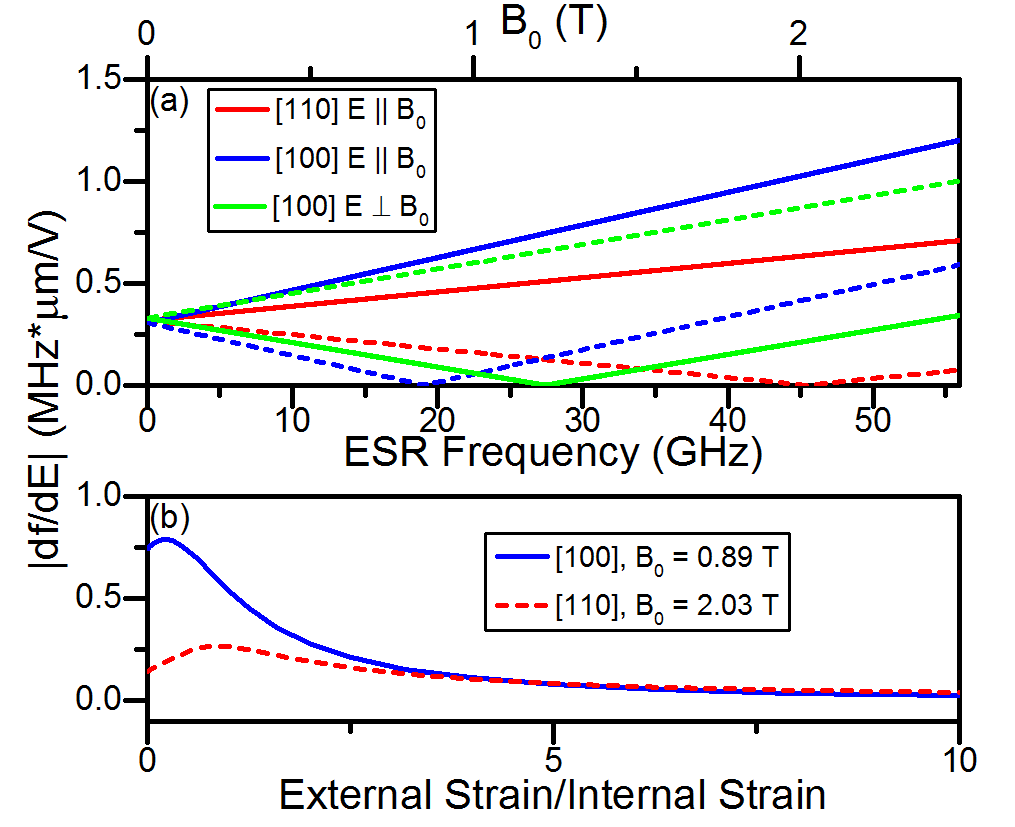}
\caption{\label{fig:fig3} (a) Plot of the electric-field sensitivity of P donors in Si for various orientations of $\vec{E}$ and $\vec{B}_{0}$ as a function of $B_{0}$. This plot assumes an internal strain of $1 \ V/\mu m$ directed along $\vec{E}$. The dashed lines indicate $M_{I}=-1/2$ whereas the solid lines indicate $M_{I}=+1/2$. $\vec{E}$ noise is suppressed at 19.12 GHz, 27.41 GHz, and 45.32 GHz. Another minima for the $[$110$]$ oriented $\vec{E} \perp \vec{B_{0}}$ occurs at 56.65 GHz but is not shown. (b) Plot of the maximum (worst case) $\vec{E}$ sensitivity of spins subject to random strain (with magnitude $1 \ V/\mu m$) as a function of externally applied strain. This plot assumes the external strain is oriented perpendicular to the $\vec{E}$ noise.} 
\end{figure}

Taking into account strain, we quantitatively compute the sensitivity of spins to electric-field noise ($df/dE$). For the simple case, where the strain is uniform and oriented in the same direction as the $\vec{E}_{noise}$, we plot $|df/dE|$ in Fig.~\ref{fig:fig3}(a) (assuming $\vec{E}_{strain}$ is $1 \ V/\mu m$). $\vec{E}$ noise suppression points appear as minima in $|df/dE|$. This case applies to bulk $^{28}$Si crystals where the strain distribution can be small ($< 10^{-3} \ V/\mu m$) and single-donor systems.

We recognize that in devices containing ensembles of spins, randomly oriented strain leads to a distribution in $\eta_{g}$ over the ensemble, washing out noise suppression effects. Assuming random strain of order $1 \ V/\mu m$, comparable to the strain in our samples, $df/dE$ should on average decrease by a factor of 5 for one hyperfine line compared to the other. This corresponds to an increase in $T_{2}$ by a factor of 25 (because $1/T_{2} \propto (df/dE)^{2}$ \cite{desousa2009}) as shown in the supplemental information. This implies that, even with the simple approach of choosing the optimal field and hyperfine line, one can substantially suppress $\vec{E}$ noise.

We propose applying a large uniform external strain or DC electric field as a remedy to random strain effects. This external field adds to $\vec{E}_{strain}$ such that the total field becomes nearly uniform. The overall $\vec{E}$ vector is pinned along the external-field axis and is insensitive to any small variations in $\vec{E}_{strain}$. External strain fields of up to 10 $V/\mu m$ have been studied \cite{feher1961} and for our samples we calculate that applying a strain of this magnitude would decrease $df/dE$ by a factor of $\sim 25$ (increases $T_{2}$ by two orders of magnitude). However, one must take care when orientating the external strain. If it is large and oriented perpendicular to the electric-field noise, a slight deviation from perfect alignment can lead to substantial decoherence due to the linear term in Eq.(3). Applying strain perpendicular to the $\vec{E}$ noise will suppress the linear term and protect the donor spin from electric-field induced decoherence. Fig.~\ref{fig:fig3}(b) shows the effect of external strain on the magnitude of $|df/dE|$.

While electric-field noise suppression points seemingly undermine the electrostatic addressability of donors, only one of the two hyperfine lines for phosphorus is protected. Global RF pulses can be used to flip the nuclear spins, toggling between electric field protected and sensitive states. Furthermore, applying an external strain or electric field enhances the Stark effect for spins in the sensitive state.

In conclusion, we have measured the quadratic Stark shift for phosphorus donors in silicon using a novel, capacitively-terminated CPW resonator. We resolved both a hyperfine and a highly anisotropic spin-orbit Stark shift. We measured $\vec{E}_{strain}$ in our samples to be on the order of $1 \ V / \mu m$ and showed that this leads to a large linear Stark shift for even small applied electric fields, making spins sensitive to electric-field noise. Using our data, we predict DC magnetic fields where electric-field noise can be suppressed. In the presence of randomly distributed internal strains, the noise suppression is weakened, but by choosing the correct ESR transition, we calculate that one can enhance $T_{2}$ by a factor of 25. We have proposed the use of large external strains to overcome this limitation such that $T_{2}$ can then be extended by two orders of magnitude. While the noise suppression techniques described in this paper use phosphorus donors in Si as an example, they should extend to other donor qubits as well.

\subsection{}
\subsubsection{}

\begin{acknowledgments}
Work was supported by the NSF through the Materials World Network Program (DMR-1107606), the ARO (W911NF-13-1-0179), and Princeton MRSEC (DMR-01420541).
\end{acknowledgments}

\providecommand{\noopsort}[1]{}\providecommand{\singleletter}[1]{#1}%
%

\widetext
\clearpage
\begin{center}
\section{\large Anisotropic Stark effect and electric-field noise suppression for phosphorus donor qubits in silicon: Supplemental Materials}
\end{center}
\setcounter{equation}{0}
\setcounter{figure}{0}
\setcounter{table}{0}
\setcounter{page}{1}
\makeatletter
\renewcommand{\theequation}{S\arabic{equation}}
\renewcommand{\thefigure}{S\arabic{figure}}
\renewcommand{\bibnumfmt}[1]{[S#1]}
\renewcommand{\citenumfont}[1]{S#1}

\section{1. \boldmath$B_{1}$ field inhomogeneity in the coplanar waveguide}

\begin{figure}[h]
\includegraphics{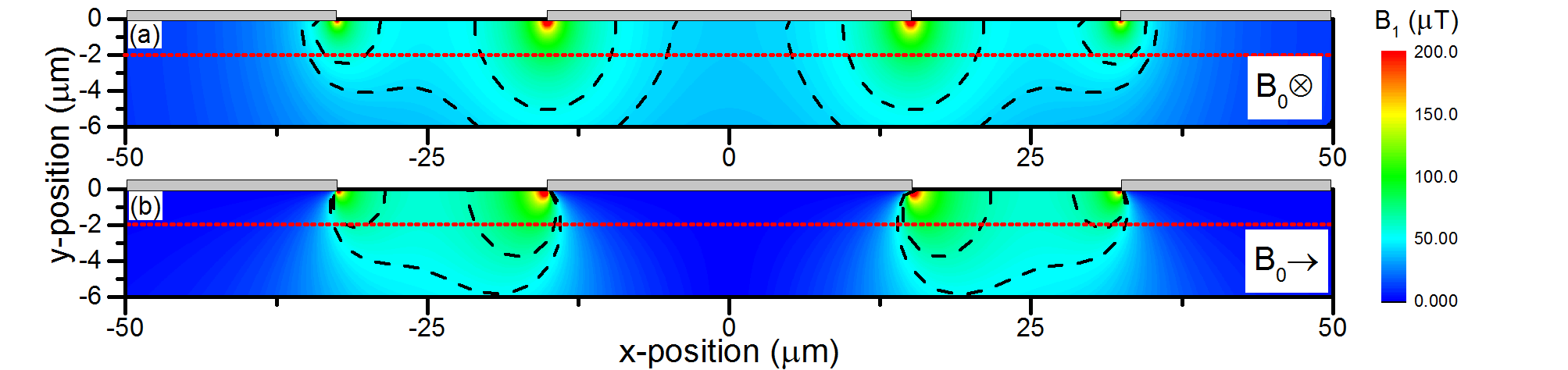}
\caption{\label{fig:SM1} Plot of CPW cross section showing the component of $\vec{B}_{1}$ driving spin rotations for (a) $\vec{E} \perp \vec{B}_{0}$ and (b)$\vec{E} \ || \ \vec{B}_{0}$ at an antinode in $\vec{B}_{1}$. The cartoons at the top of each plot represent Nb conductors making up the waveguide. The regions for which spins should contribute to the signal (light blue) are bounded by the black dashed lines. The red dotted line, $2 \ \mu m$ below the CPW, indicates the boundary of the spin ensemble so the signal only originates from areas above this line.} 
\end{figure}

As discussed in the main text, only the component of $\vec{B}_{1}$ perpendicular to $\vec{B}_{0}$ drives spin rotations. Depending on the sample orientation in $B_{0}$, this $\vec{B}_{1}$ component can change. In Fig. \ref{fig:SM1} we plot the component of $\vec{B}_{1}$ perpendicular to $\vec{B_{0}}$ for both orientations discussed in the main text. In these experiments, microwave power was tuned to enhance sensitivity to spins near the center of the gap where $\vec{E}$ is most uniform. The typical microwave power used was $\sim 2 \ \mu W$ for $400 \ ns$ $\pi$-pulses.

\section{2. Calculation of strain distribution from unipolar pulse data}

\begin{figure}[h]
\includegraphics{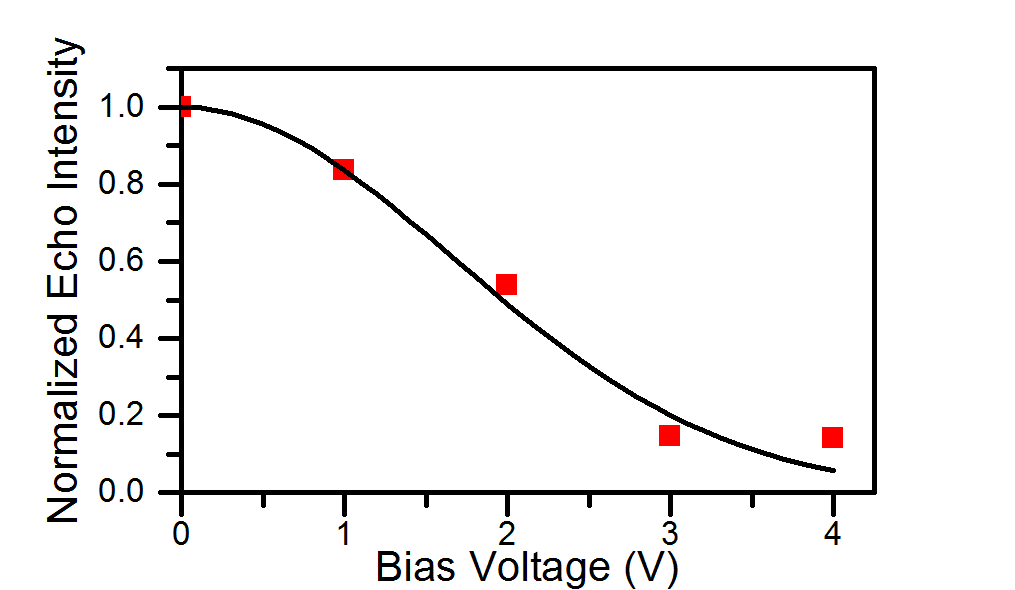}
\caption{\label{fig:figSM2} Plot of the normalized ESR signal intensity as unipolar pulse voltage is varied. The decrease in the signal indicates a distribution in the linear Stark effect which leads to additional dephasing in the spin ensemble. The data is fitted using Eq.(2) and (3) from the main text. The best fit is obtained when assuming a Gaussian distribution of strain with a standard deviation of $0.8 \ V/\mu m$. } 
\end{figure}

When applying unipolar electric-field pulses, in addition to the Hahn echo phase shift, we observed a rapid decrease of the ESR signal as the bias amplitude was increased. This is due to a distribution in the Stark shifts over the spin ensemble. Recovery of the echo signal when applying bipolar electric-field pulses indicate that the distribution is due to linear Stark effects. The unipolar pulse data taken with bias amplitudes up to $4 \ V$ are displayed in Fig.~\ref{fig:figSM2}. 

We model the signal loss due to line broadening from unipolar electric-field pulses by putting the Stark-induced phase shift from Eq.(3) into the model described by Eq.(2) in the main text. By assuming a Gaussian distribution of $E_{strain}$ we fit to the observed decrease in echo amplitude as a function of applied bias voltage as shown in Fig.~\ref{fig:figSM2}. We find that the strain distribution has a standard deviation of $0.8 \ V/\mu m$.

\section{3. Calculation of electric-field noise suppression}

To quantify the effect of electric-field noise on $T_{2}$, we have modelled the decoherence for our devices in the Bloch-Wangness-Redfield limit using reference [22] of the main text. Assuming realistic values for internal strain and assuming that the electric-field noise comes from voltage fluctuations on the gate, we plot the $T_{2}$ decays for spins both at and away from the noise suppression point. 
\subsection{}
\subsubsection{}

\begin{figure}[h]
\includegraphics{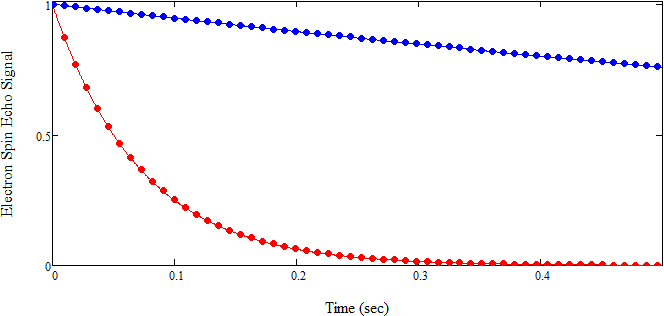}
\caption{\label{fig:figSM3}Electric-field noise effect on electron spin decoherence of $^{31}$P donors in silicon. The T$_2$ decays (solid dots) were simulated for an ESR transition far from the Stark-effect cancellation point (red) and exactly at the cancellation point (blue). The spins were assumed to be subject to a random internal stress of 1~V/$\mu$m, and a white noise spectrum was assumed with a 10~$\mu$V magnitude on the central pin of the CPW resonator. Solid lines are exponential fits revealing a $T_{2}$ of 72~ms for the red curve and 1.8~s for the blue curve. } 
\end{figure}


\begin{thebibliography}{22}%
\makeatletter
\providecommand \@ifxundefined [1]{%
 \@ifx{#1\undefined}
}%
\providecommand \@ifnum [1]{%
 \ifnum #1\expandafter \@firstoftwo
 \else \expandafter \@secondoftwo
 \fi
}%
\providecommand \@ifx [1]{%
 \ifx #1\expandafter \@firstoftwo
 \else \expandafter \@secondoftwo
 \fi
}%
\providecommand \natexlab [1]{#1}%
\providecommand \enquote  [1]{``#1''}%
\providecommand \bibnamefont  [1]{#1}%
\providecommand \bibfnamefont [1]{#1}%
\providecommand \citenamefont [1]{#1}%
\providecommand \href@noop [0]{\@secondoftwo}%
\providecommand \href [0]{\begingroup \@sanitize@url \@href}%
\providecommand \@href[1]{\@@startlink{#1}\@@href}%
\providecommand \@@href[1]{\endgroup#1\@@endlink}%
\providecommand \@sanitize@url [0]{\catcode `\\12\catcode `\$12\catcode
  `\&12\catcode `\#12\catcode `\^12\catcode `\_12\catcode `\%12\relax}%
\providecommand \@@startlink[1]{}%
\providecommand \@@endlink[0]{}%
\providecommand \url  [0]{\begingroup\@sanitize@url \@url }%
\providecommand \@url [1]{\endgroup\@href {#1}{\urlprefix }}%
\providecommand \urlprefix  [0]{URL }%
\providecommand \Eprint [0]{\href }%
\providecommand \doibase [0]{http://dx.doi.org/}%
\providecommand \selectlanguage [0]{\@gobble}%
\providecommand \bibinfo  [0]{\@secondoftwo}%
\providecommand \bibfield  [0]{\@secondoftwo}%
\providecommand \translation [1]{[#1]}%
\providecommand \BibitemOpen [0]{}%
\providecommand \bibitemStop [0]{}%
\providecommand \bibitemNoStop [0]{.\EOS\space}%
\providecommand \EOS [0]{\spacefactor3000\relax}%
\providecommand \BibitemShut  [1]{\csname bibitem#1\endcsname}%
\let\auto@bib@innerbib\@empty
\bibitem [{\citenamefont {Kane}(1998)}]{kane1998}%
  \BibitemOpen
  \bibfield  {author} {\bibinfo {author} {\bibfnamefont {B.~E.}\ \bibnamefont
  {Kane}},\ }\href {\doibase 10.1038/30156} {\bibfield  {journal} {\bibinfo
  {journal} {Nature}\ }\textbf {\bibinfo {volume} {393}},\ \bibinfo {pages}
  {133} (\bibinfo {year} {1998})}\BibitemShut {NoStop}%
\bibitem [{\citenamefont {Morton}\ \emph {et~al.}(2011)\citenamefont {Morton},
  \citenamefont {McCamey}, \citenamefont {Eriksson},\ and\ \citenamefont
  {Lyon}}]{morton2011}%
  \BibitemOpen
  \bibfield  {author} {\bibinfo {author} {\bibfnamefont {J.~J.~L.}\
  \bibnamefont {Morton}}, \bibinfo {author} {\bibfnamefont {D.~R.}\
  \bibnamefont {McCamey}}, \bibinfo {author} {\bibfnamefont {M.~A.}\
  \bibnamefont {Eriksson}}, \ and\ \bibinfo {author} {\bibfnamefont {S.~A.}\
  \bibnamefont {Lyon}},\ }\href {\doibase 10.1038/nature10681} {\bibfield
  {journal} {\bibinfo  {journal} {Nature}\ }\textbf {\bibinfo {volume} {479}},\
  \bibinfo {pages} {345} (\bibinfo {year} {2011})}\BibitemShut {NoStop}%
\bibitem [{\citenamefont {Hill}\ \emph {et~al.}(2005)\citenamefont {Hill},
  \citenamefont {Hollenberg}, \citenamefont {Fowler}, \citenamefont {Wellard},
  \citenamefont {Greentree},\ and\ \citenamefont {Goan}}]{hill2005}%
  \BibitemOpen
  \bibfield  {author} {\bibinfo {author} {\bibfnamefont {C.~D.}\ \bibnamefont
  {Hill}}, \bibinfo {author} {\bibfnamefont {L.~C.~L.}\ \bibnamefont
  {Hollenberg}}, \bibinfo {author} {\bibfnamefont {A.~G.}\ \bibnamefont
  {Fowler}}, \bibinfo {author} {\bibfnamefont {C.~J.}\ \bibnamefont {Wellard}},
  \bibinfo {author} {\bibfnamefont {A.~D.}\ \bibnamefont {Greentree}}, \ and\
  \bibinfo {author} {\bibfnamefont {H.-S.}\ \bibnamefont {Goan}},\ }\href
  {\doibase 10.1103/PhysRevB.72.045350} {\bibfield  {journal} {\bibinfo
  {journal} {Phys. Rev. B}\ }\textbf {\bibinfo {volume} {72}},\ \bibinfo
  {pages} {045350} (\bibinfo {year} {2005})}\BibitemShut {NoStop}%
\bibitem [{\citenamefont {Tyryshkin}\ \emph {et~al.}(2012)\citenamefont
  {Tyryshkin}, \citenamefont {Tojo}, \citenamefont {Morton}, \citenamefont
  {Riemann}, \citenamefont {Abrosimov}, \citenamefont {Becker}, \citenamefont
  {Pohl}, \citenamefont {Schenkel}, \citenamefont {Thewalt}, \citenamefont
  {Itoh},\ and\ \citenamefont {Lyon}}]{tyryshkin2012}%
  \BibitemOpen
  \bibfield  {author} {\bibinfo {author} {\bibfnamefont {A.~M.}\ \bibnamefont
  {Tyryshkin}}, \bibinfo {author} {\bibfnamefont {S.}~\bibnamefont {Tojo}},
  \bibinfo {author} {\bibfnamefont {J.~J.~L.}\ \bibnamefont {Morton}}, \bibinfo
  {author} {\bibfnamefont {H.}~\bibnamefont {Riemann}}, \bibinfo {author}
  {\bibfnamefont {N.~V.}\ \bibnamefont {Abrosimov}}, \bibinfo {author}
  {\bibfnamefont {P.}~\bibnamefont {Becker}}, \bibinfo {author} {\bibfnamefont
  {H.-J.}\ \bibnamefont {Pohl}}, \bibinfo {author} {\bibfnamefont
  {T.}~\bibnamefont {Schenkel}}, \bibinfo {author} {\bibfnamefont {M.~L.~W.}\
  \bibnamefont {Thewalt}}, \bibinfo {author} {\bibfnamefont {K.~M.}\
  \bibnamefont {Itoh}}, \ and\ \bibinfo {author} {\bibfnamefont {S.~A.}\
  \bibnamefont {Lyon}},\ }\href {\doibase 10.1038/nmat3182} {\bibfield
  {journal} {\bibinfo  {journal} {Nat Mater}\ }\textbf {\bibinfo {volume}
  {11}},\ \bibinfo {pages} {143} (\bibinfo {year} {2012})}\BibitemShut
  {NoStop}%
\bibitem [{\citenamefont {Wolfowicz}\ \emph {et~al.}(2013)\citenamefont
  {Wolfowicz}, \citenamefont {Tyryshkin}, \citenamefont {George}, \citenamefont
  {Riemann}, \citenamefont {Abrosimov}, \citenamefont {Becker}, \citenamefont
  {Pohl}, \citenamefont {Thewalt}, \citenamefont {Lyon},\ and\ \citenamefont
  {Morton}}]{wolfowicz2013}%
  \BibitemOpen
  \bibfield  {author} {\bibinfo {author} {\bibfnamefont {G.}~\bibnamefont
  {Wolfowicz}}, \bibinfo {author} {\bibfnamefont {A.~M.}\ \bibnamefont
  {Tyryshkin}}, \bibinfo {author} {\bibfnamefont {R.~E.}\ \bibnamefont
  {George}}, \bibinfo {author} {\bibfnamefont {H.}~\bibnamefont {Riemann}},
  \bibinfo {author} {\bibfnamefont {N.~V.}\ \bibnamefont {Abrosimov}}, \bibinfo
  {author} {\bibfnamefont {P.}~\bibnamefont {Becker}}, \bibinfo {author}
  {\bibfnamefont {H.-J.}\ \bibnamefont {Pohl}}, \bibinfo {author}
  {\bibfnamefont {M.~L.~W.}\ \bibnamefont {Thewalt}}, \bibinfo {author}
  {\bibfnamefont {S.~A.}\ \bibnamefont {Lyon}}, \ and\ \bibinfo {author}
  {\bibfnamefont {J.~J.~L.}\ \bibnamefont {Morton}},\ }\href {\doibase
  10.1038/nnano.2013.117} {\bibfield  {journal} {\bibinfo  {journal} {Nat
  Nano}\ }\textbf {\bibinfo {volume} {8}},\ \bibinfo {pages} {561} (\bibinfo
  {year} {2013})}\BibitemShut {NoStop}%
\bibitem [{\citenamefont {de~Sousa}\ \emph {et~al.}(2004)\citenamefont
  {de~Sousa}, \citenamefont {Delgado},\ and\ \citenamefont
  {Das~Sarma}}]{desousa2004}%
  \BibitemOpen
  \bibfield  {author} {\bibinfo {author} {\bibfnamefont {R.}~\bibnamefont
  {de~Sousa}}, \bibinfo {author} {\bibfnamefont {J.~D.}\ \bibnamefont
  {Delgado}}, \ and\ \bibinfo {author} {\bibfnamefont {S.}~\bibnamefont
  {Das~Sarma}},\ }\href {\doibase 10.1103/PhysRevA.70.052304} {\bibfield
  {journal} {\bibinfo  {journal} {Phys. Rev. A}\ }\textbf {\bibinfo {volume}
  {70}},\ \bibinfo {pages} {052304} (\bibinfo {year} {2004})}\BibitemShut
  {NoStop}%
\bibitem [{\citenamefont {Hollenberg}\ \emph {et~al.}(2006)\citenamefont
  {Hollenberg}, \citenamefont {Greentree}, \citenamefont {Fowler},\ and\
  \citenamefont {Wellard}}]{hollenberg2006}%
  \BibitemOpen
  \bibfield  {author} {\bibinfo {author} {\bibfnamefont {L.~C.~L.}\
  \bibnamefont {Hollenberg}}, \bibinfo {author} {\bibfnamefont {A.~D.}\
  \bibnamefont {Greentree}}, \bibinfo {author} {\bibfnamefont {A.~G.}\
  \bibnamefont {Fowler}}, \ and\ \bibinfo {author} {\bibfnamefont {C.~J.}\
  \bibnamefont {Wellard}},\ }\href {\doibase 10.1103/PhysRevB.74.045311}
  {\bibfield  {journal} {\bibinfo  {journal} {Phys. Rev. B}\ }\textbf {\bibinfo
  {volume} {74}},\ \bibinfo {pages} {045311} (\bibinfo {year}
  {2006})}\BibitemShut {NoStop}%
\bibitem [{\citenamefont {Bradbury}\ \emph {et~al.}(2006)\citenamefont
  {Bradbury}, \citenamefont {Tyryshkin}, \citenamefont {Sabouret},
  \citenamefont {Bokor}, \citenamefont {Schenkel},\ and\ \citenamefont
  {Lyon}}]{bradbury2006}%
  \BibitemOpen
  \bibfield  {author} {\bibinfo {author} {\bibfnamefont {F.~R.}\ \bibnamefont
  {Bradbury}}, \bibinfo {author} {\bibfnamefont {A.~M.}\ \bibnamefont
  {Tyryshkin}}, \bibinfo {author} {\bibfnamefont {G.}~\bibnamefont {Sabouret}},
  \bibinfo {author} {\bibfnamefont {J.}~\bibnamefont {Bokor}}, \bibinfo
  {author} {\bibfnamefont {T.}~\bibnamefont {Schenkel}}, \ and\ \bibinfo
  {author} {\bibfnamefont {S.~A.}\ \bibnamefont {Lyon}},\ }\href {\doibase
  10.1103/PhysRevLett.97.176404} {\bibfield  {journal} {\bibinfo  {journal}
  {Phys. Rev. Lett.}\ }\textbf {\bibinfo {volume} {97}},\ \bibinfo {pages}
  {176404} (\bibinfo {year} {2006})}\BibitemShut {NoStop}%
\bibitem [{\citenamefont {Lo}\ \emph {et~al.}(2014)\citenamefont {Lo},
  \citenamefont {Simmons}, \citenamefont {Lo~Nardo}, \citenamefont {Weis},
  \citenamefont {Tyryshkin}, \citenamefont {Meijer}, \citenamefont {Rogalla},
  \citenamefont {Lyon}, \citenamefont {Bokor}, \citenamefont {Schenkel},\ and\
  \citenamefont {Morton}}]{lo2014}%
  \BibitemOpen
  \bibfield  {author} {\bibinfo {author} {\bibfnamefont {C.~C.}\ \bibnamefont
  {Lo}}, \bibinfo {author} {\bibfnamefont {S.}~\bibnamefont {Simmons}},
  \bibinfo {author} {\bibfnamefont {R.}~\bibnamefont {Lo~Nardo}}, \bibinfo
  {author} {\bibfnamefont {C.~D.}\ \bibnamefont {Weis}}, \bibinfo {author}
  {\bibfnamefont {A.~M.}\ \bibnamefont {Tyryshkin}}, \bibinfo {author}
  {\bibfnamefont {J.}~\bibnamefont {Meijer}}, \bibinfo {author} {\bibfnamefont
  {D.}~\bibnamefont {Rogalla}}, \bibinfo {author} {\bibfnamefont {S.~A.}\
  \bibnamefont {Lyon}}, \bibinfo {author} {\bibfnamefont {J.}~\bibnamefont
  {Bokor}}, \bibinfo {author} {\bibfnamefont {T.}~\bibnamefont {Schenkel}}, \
  and\ \bibinfo {author} {\bibfnamefont {J.~J.~L.}\ \bibnamefont {Morton}},\
  }\href {\doibase http://dx.doi.org/10.1063/1.4876175} {\bibfield  {journal}
  {\bibinfo  {journal} {Appl. Phys. Lett.}\ }\textbf {\bibinfo {volume}
  {104}},\ \bibinfo {eid} {193502} (\bibinfo {year} {2014})}\BibitemShut
  {NoStop}%
\bibitem [{\citenamefont {Mims}(1974)}]{mims1974}%
  \BibitemOpen
  \bibfield  {author} {\bibinfo {author} {\bibfnamefont {W.~B.}\ \bibnamefont
  {Mims}},\ }\href {\doibase http://dx.doi.org/10.1063/1.1686567} {\bibfield
  {journal} {\bibinfo  {journal} {Review of Scientific Instruments}\ }\textbf
  {\bibinfo {volume} {45}},\ \bibinfo {pages} {1583} (\bibinfo {year}
  {1974})}\BibitemShut {NoStop}%
\bibitem [{\citenamefont {Dreher}\ \emph {et~al.}(2011)\citenamefont {Dreher},
  \citenamefont {Hilker}, \citenamefont {Brandlmaier}, \citenamefont
  {Goennenwein}, \citenamefont {Huebl}, \citenamefont {Stutzmann},\ and\
  \citenamefont {Brandt}}]{dreher2011}%
  \BibitemOpen
  \bibfield  {author} {\bibinfo {author} {\bibfnamefont {L.}~\bibnamefont
  {Dreher}}, \bibinfo {author} {\bibfnamefont {T.~A.}\ \bibnamefont {Hilker}},
  \bibinfo {author} {\bibfnamefont {A.}~\bibnamefont {Brandlmaier}}, \bibinfo
  {author} {\bibfnamefont {S.~T.~B.}\ \bibnamefont {Goennenwein}}, \bibinfo
  {author} {\bibfnamefont {H.}~\bibnamefont {Huebl}}, \bibinfo {author}
  {\bibfnamefont {M.}~\bibnamefont {Stutzmann}}, \ and\ \bibinfo {author}
  {\bibfnamefont {M.~S.}\ \bibnamefont {Brandt}},\ }\href {\doibase
  10.1103/PhysRevLett.106.037601} {\bibfield  {journal} {\bibinfo  {journal}
  {Phys. Rev. Lett.}\ }\textbf {\bibinfo {volume} {106}},\ \bibinfo {pages}
  {037601} (\bibinfo {year} {2011})}\BibitemShut {NoStop}%
\bibitem [{\citenamefont {Rahman}\ \emph {et~al.}(2009)\citenamefont {Rahman},
  \citenamefont {Park}, \citenamefont {Boykin}, \citenamefont {Klimeck},
  \citenamefont {Rogge},\ and\ \citenamefont {Hollenberg}}]{rahman2009}%
  \BibitemOpen
  \bibfield  {author} {\bibinfo {author} {\bibfnamefont {R.}~\bibnamefont
  {Rahman}}, \bibinfo {author} {\bibfnamefont {S.~H.}\ \bibnamefont {Park}},
  \bibinfo {author} {\bibfnamefont {T.~B.}\ \bibnamefont {Boykin}}, \bibinfo
  {author} {\bibfnamefont {G.}~\bibnamefont {Klimeck}}, \bibinfo {author}
  {\bibfnamefont {S.}~\bibnamefont {Rogge}}, \ and\ \bibinfo {author}
  {\bibfnamefont {L.~C.~L.}\ \bibnamefont {Hollenberg}},\ }\href {\doibase
  10.1103/PhysRevB.80.155301} {\bibfield  {journal} {\bibinfo  {journal} {Phys.
  Rev. B}\ }\textbf {\bibinfo {volume} {80}},\ \bibinfo {pages} {155301}
  (\bibinfo {year} {2009})}\BibitemShut {NoStop}%
\bibitem [{\citenamefont {Pica}\ \emph {et~al.}(2014)\citenamefont {Pica},
  \citenamefont {Wolfowicz}, \citenamefont {Urdampilleta}, \citenamefont
  {Thewalt}, \citenamefont {Riemann}, \citenamefont {Abrosimov}, \citenamefont
  {Becker}, \citenamefont {Pohl}, \citenamefont {Morton}, \citenamefont
  {Bhatt}, \citenamefont {Lyon},\ and\ \citenamefont {Lovett}}]{pica2014}%
  \BibitemOpen
  \bibfield  {author} {\bibinfo {author} {\bibfnamefont {G.}~\bibnamefont
  {Pica}}, \bibinfo {author} {\bibfnamefont {G.}~\bibnamefont {Wolfowicz}},
  \bibinfo {author} {\bibfnamefont {M.}~\bibnamefont {Urdampilleta}}, \bibinfo
  {author} {\bibfnamefont {M.~L.~W.}\ \bibnamefont {Thewalt}}, \bibinfo
  {author} {\bibfnamefont {H.}~\bibnamefont {Riemann}}, \bibinfo {author}
  {\bibfnamefont {N.~V.}\ \bibnamefont {Abrosimov}}, \bibinfo {author}
  {\bibfnamefont {P.}~\bibnamefont {Becker}}, \bibinfo {author} {\bibfnamefont
  {H.-J.}\ \bibnamefont {Pohl}}, \bibinfo {author} {\bibfnamefont {J.~J.~L.}\
  \bibnamefont {Morton}}, \bibinfo {author} {\bibfnamefont {R.~N.}\
  \bibnamefont {Bhatt}}, \bibinfo {author} {\bibfnamefont {S.~A.}\ \bibnamefont
  {Lyon}}, \ and\ \bibinfo {author} {\bibfnamefont {B.~W.}\ \bibnamefont
  {Lovett}},\ }\href {\doibase 10.1103/PhysRevB.90.195204} {\bibfield
  {journal} {\bibinfo  {journal} {Phys. Rev. B}\ }\textbf {\bibinfo {volume}
  {90}},\ \bibinfo {pages} {195204} (\bibinfo {year} {2014})}\BibitemShut
  {NoStop}%
\bibitem [{\citenamefont {Simons}(2001)}]{simons2001}%
  \BibitemOpen
  \bibfield  {author} {\bibinfo {author} {\bibfnamefont {R.~N.}\ \bibnamefont
  {Simons}},\ }\href@noop {} {\emph {\bibinfo {title} {Coplanar Waveguide
  Circuits, Components, and Systems}}}\ (\bibinfo  {publisher} {Wiley
  Interscience},\ \bibinfo {year} {2001})\BibitemShut {NoStop}%
\bibitem [{\citenamefont {Wen}(1969)}]{wen1969}%
  \BibitemOpen
  \bibfield  {author} {\bibinfo {author} {\bibfnamefont {C.}~\bibnamefont
  {Wen}},\ }\href {\doibase 10.1109/TMTT.1969.1127105} {\bibfield  {journal}
  {\bibinfo  {journal} {Microwave Theory and Techniques, IEEE Transactions on}\
  }\textbf {\bibinfo {volume} {17}},\ \bibinfo {pages} {1087} (\bibinfo {year}
  {1969})}\BibitemShut {NoStop}%
\bibitem [{\citenamefont {Sigillito}\ \emph {et~al.}(2014)\citenamefont
  {Sigillito}, \citenamefont {Malissa}, \citenamefont {Tyryshkin},
  \citenamefont {Riemann}, \citenamefont {Abrosimov}, \citenamefont {Becker},
  \citenamefont {Pohl}, \citenamefont {Thewalt}, \citenamefont {Itoh},
  \citenamefont {Morton}, \citenamefont {Houck}, \citenamefont {Schuster},\
  and\ \citenamefont {Lyon}}]{sigillito2014}%
  \BibitemOpen
  \bibfield  {author} {\bibinfo {author} {\bibfnamefont {A.~J.}\ \bibnamefont
  {Sigillito}}, \bibinfo {author} {\bibfnamefont {H.}~\bibnamefont {Malissa}},
  \bibinfo {author} {\bibfnamefont {A.~M.}\ \bibnamefont {Tyryshkin}}, \bibinfo
  {author} {\bibfnamefont {H.}~\bibnamefont {Riemann}}, \bibinfo {author}
  {\bibfnamefont {N.~V.}\ \bibnamefont {Abrosimov}}, \bibinfo {author}
  {\bibfnamefont {P.}~\bibnamefont {Becker}}, \bibinfo {author} {\bibfnamefont
  {H.-J.}\ \bibnamefont {Pohl}}, \bibinfo {author} {\bibfnamefont {M.~L.~W.}\
  \bibnamefont {Thewalt}}, \bibinfo {author} {\bibfnamefont {K.~M.}\
  \bibnamefont {Itoh}}, \bibinfo {author} {\bibfnamefont {J.~J.~L.}\
  \bibnamefont {Morton}}, \bibinfo {author} {\bibfnamefont {A.~A.}\
  \bibnamefont {Houck}}, \bibinfo {author} {\bibfnamefont {D.~I.}\ \bibnamefont
  {Schuster}}, \ and\ \bibinfo {author} {\bibfnamefont {S.~A.}\ \bibnamefont
  {Lyon}},\ }\href {\doibase http://dx.doi.org/10.1063/1.4881613} {\bibfield
  {journal} {\bibinfo  {journal} {Appl. Phys. Lett.}\ }\textbf {\bibinfo
  {volume} {104}},\ \bibinfo {eid} {222407} (\bibinfo {year}
  {2014})}\BibitemShut {NoStop}%
\bibitem [{\citenamefont {Malissa}\ \emph {et~al.}(2013)\citenamefont
  {Malissa}, \citenamefont {Schuster}, \citenamefont {Tyryshkin}, \citenamefont
  {Houck},\ and\ \citenamefont {Lyon}}]{malissa2013}%
  \BibitemOpen
  \bibfield  {author} {\bibinfo {author} {\bibfnamefont {H.}~\bibnamefont
  {Malissa}}, \bibinfo {author} {\bibfnamefont {D.~I.}\ \bibnamefont
  {Schuster}}, \bibinfo {author} {\bibfnamefont {A.~M.}\ \bibnamefont
  {Tyryshkin}}, \bibinfo {author} {\bibfnamefont {A.~A.}\ \bibnamefont
  {Houck}}, \ and\ \bibinfo {author} {\bibfnamefont {S.~A.}\ \bibnamefont
  {Lyon}},\ }\href {\doibase http://dx.doi.org/10.1063/1.4792205} {\bibfield
  {journal} {\bibinfo  {journal} {Review of Scientific Instruments}\ }\textbf
  {\bibinfo {volume} {84}},\ \bibinfo {eid} {025116} (\bibinfo {year}
  {2013})}\BibitemShut {NoStop}%
\bibitem [{\citenamefont {Schweiger}\ and\ \citenamefont
  {Jeschke}(2001)}]{Schweiger2001}%
  \BibitemOpen
  \bibfield  {author} {\bibinfo {author} {\bibfnamefont {A.}~\bibnamefont
  {Schweiger}}\ and\ \bibinfo {author} {\bibfnamefont {G.}~\bibnamefont
  {Jeschke}},\ }\href@noop {} {\emph {\bibinfo {title} {Principles of pulse
  electron paramagnetic resonance}}}\ (\bibinfo  {publisher} {Oxford University
  Press},\ \bibinfo {year} {2001})\ \bibinfo {note} {section 7.2}\BibitemShut
  {NoStop}%
\bibitem [{\citenamefont {Wilson}\ and\ \citenamefont
  {Feher}(1961)}]{feher1961}%
  \BibitemOpen
  \bibfield  {author} {\bibinfo {author} {\bibfnamefont {D.~K.}\ \bibnamefont
  {Wilson}}\ and\ \bibinfo {author} {\bibfnamefont {G.}~\bibnamefont {Feher}},\
  }\href {\doibase 10.1103/PhysRev.124.1068} {\bibfield  {journal} {\bibinfo
  {journal} {Phys. Rev.}\ }\textbf {\bibinfo {volume} {124}},\ \bibinfo {pages}
  {1068} (\bibinfo {year} {1961})}\BibitemShut {NoStop}%
\bibitem [{\citenamefont {Bradbury}\ \emph {et~al.}(2007)\citenamefont
  {Bradbury}, \citenamefont {Tyryshkin}, \citenamefont {Sabouret},
  \citenamefont {Bokor}, \citenamefont {Schenkel},\ and\ \citenamefont
  {Lyon}}]{bradbury2007aip}%
  \BibitemOpen
  \bibfield  {author} {\bibinfo {author} {\bibfnamefont {F.~R.}\ \bibnamefont
  {Bradbury}}, \bibinfo {author} {\bibfnamefont {A.~M.}\ \bibnamefont
  {Tyryshkin}}, \bibinfo {author} {\bibfnamefont {G.}~\bibnamefont {Sabouret}},
  \bibinfo {author} {\bibfnamefont {J.}~\bibnamefont {Bokor}}, \bibinfo
  {author} {\bibfnamefont {T.}~\bibnamefont {Schenkel}}, \ and\ \bibinfo
  {author} {\bibfnamefont {S.~A.}\ \bibnamefont {Lyon}},\ }\href {\doibase
  10.1063/1.2730277} {\bibfield  {journal} {\bibinfo  {journal} {AIP Conf.
  Proc.}\ }\textbf {\bibinfo {volume} {893}},\ \bibinfo {pages} {1093}
  (\bibinfo {year} {2007})}\BibitemShut {NoStop}%
\bibitem [{\citenamefont {Tyryshkin}\ \emph {et~al.}(2006)\citenamefont
  {Tyryshkin}, \citenamefont {Morton}, \citenamefont {Ardavan},\ and\
  \citenamefont {Lyon}}]{tyryshkin2006}%
  \BibitemOpen
  \bibfield  {author} {\bibinfo {author} {\bibfnamefont {A.~M.}\ \bibnamefont
  {Tyryshkin}}, \bibinfo {author} {\bibfnamefont {J.~J.~L.}\ \bibnamefont
  {Morton}}, \bibinfo {author} {\bibfnamefont {A.}~\bibnamefont {Ardavan}}, \
  and\ \bibinfo {author} {\bibfnamefont {S.~A.}\ \bibnamefont {Lyon}},\ }\href
  {\doibase http://dx.doi.org/10.1063/1.2204915} {\bibfield  {journal}
  {\bibinfo  {journal} {J. Chem. Phys.}\ }\textbf {\bibinfo {volume} {124}},\
  \bibinfo {eid} {234508} (\bibinfo {year} {2006})}\BibitemShut {NoStop}%
\bibitem [{\citenamefont {de~Sousa}(2009)}]{desousa2009}%
  \BibitemOpen
  \bibfield  {author} {\bibinfo {author} {\bibfnamefont {R.}~\bibnamefont
  {de~Sousa}},\ }in\ \href {\doibase 10.1007/978-3-540-79365-6_10} {\emph
  {\bibinfo {booktitle} {Electron Spin Resonance and Related Phenomena in
  Low-Dimensional Structures}}},\ \bibinfo {series} {Topics in Applied
  Physics}, Vol.\ \bibinfo {volume} {115},\ \bibinfo {editor} {edited by\
  \bibinfo {editor} {\bibfnamefont {M.}~\bibnamefont {Fanciulli}}}\ (\bibinfo
  {publisher} {Springer Berlin Heidelberg},\ \bibinfo {year} {2009})\ pp.\
  \bibinfo {pages} {183--220}\BibitemShut {NoStop}%
\end{thebibliography}
\end{document}